\documentclass[letterpaper,floatfix,aps,prl,amsmath,amssymb,longbibliography,twocolumn,showpacs]{revtex4}
\usepackage{hyperref}
\usepackage{color}
\usepackage{comment}
\usepackage{amsmath,amssymb,graphicx,braket,dsfont}
\DeclareMathOperator{\Tr}{Tr}
\DeclareMathOperator{\arctanh}{arctanh}
\usepackage[latin1]{inputenc}
\usepackage[caption=false]{subfig}

\begin{document}

\title{Dicke simulators with emergent collective quantum computational abilities}
\author{Pietro Rotondo$^1$, Marco Cosentino Lagomarsino$^{2}$, and Giovanni Viola$^{3,4}$ }
\affiliation{$^{1}$ Dipartimento di Fisica, Universit\`a degli Studi di Milano and INFN, via Celoria 16, 20133 Milano, Italy\\
$^{2}$Sorbonne Universit\'{e}s and CNRS, UPMC Univ Paris 06, UMR 7238, Computational and Quantitative Biology, 15 Rue de l'\'{E}cole de M\'{e}decine, Paris, France\\
$^{3}$ Department of Microtechnology and Nanoscience (MC2),
Chalmers University of Technology, SE-412 96 Gothenburg, Sweden\\
$^{4}$ Institute for Quantum Information, RWTH Aachen University, D-52056 Aachen, Germany}
\date{\today}

\begin{abstract}
  Using an approach inspired from Spin Glasses, we show that the
  multimode disordered Dicke model is equivalent to a quantum Hopfield
  network. 
  We propose variational ground states for the system
  at zero temperature, which we conjecture to be exact in the
  thermodynamic limit. These ground states contain the information on
  the disordered qubit-photon couplings. 
  These results lead to two intriguing physical implications. First,
  once the qubit-photon couplings can be engineered, it should be
  possible to build scalable pattern-storing systems whose dynamics is governed
  by quantum laws. Second, we argue with an example how such Dicke
  quantum simulators might be used as a solver of ``hard''
  combinatorial optimization problems.
\end{abstract}

\maketitle

The connection of experimentally realizable quantum systems with
  computation contains promising perspectives from both the
  fundamental and the technological viewpoint~\cite{Feynman82,nielsen2010quantum}.  For example,
quantum computational capabilities can be implemented by ``quantum gates''~\cite{PhysRevA.52.3457}
and by the so-called ``adiabatic quantum optimization''
technique~\cite{Farhi20042001,Santoro29032002,Zamponi:12}.
Today's experimental technology of highly controllable
quantum simulators, recently used  for testing theoretical predictions in
a wide range of areas of
physics~\cite{Buluta02102009,RevModPhys.86.153,lewenstein2007ultracold}, offers new opportunities
for exploring computing power for quantum systems.

In the case of light-matter interaction at the quantum level,
the reference benchmark is the Dicke
  model~\cite{Dicke:PhysRev:54}. Studies of its equilibrium
properties have predicted a superradiant transition to occur in the
strong coupling and low temperature
regime~\cite{Lieb:AnnPhys:73,Lieb:PRA:73,Wang:PRA:73}. The
superradiant phase is characterized by a macroscopic number of atoms
in the excited state whose collective behaviour produces an
enhancement of spontaneous emission (proportional to the number of
cooperating atoms in the sample). Crucially, this phenomenology is in
direct link with experimentally feasible quantum simulators.  Recently, Nagy and
coworkers~\cite{Nagy:PRL:10} argued that the Dicke model effectively
describes the self-organization phase transition of a Bose-Einstein
condensate (BEC) in an optical
cavity~\cite{Baumann:Nat:10,Baumann:PRL:11}.  Additionally, Dimer and
colleagues~\cite{Dimer:PRA:07} proposed a Cavity QED realization of
the Dicke model based on cavity-mediated Raman transitions, closer in spirit to
the original Dicke's idea. Evidence of superradiance in this system is
reported in~\cite{Baden:PRL:14}. An implementation of generalized
Dicke models in hybrid quantum systems has also been put
forward~\cite{Zou:PRL:14}. More generally, Dicke-like Hamiltonians
describe a variety of physical systems, ranging from Circuit QED
\cite{Nori:SciRep:14,Nataf:NatComm:10,Viehmann:PRL:11,Mlinek:NatComm:14,Mezzacapo:SciRep:14}
to Cavity QED with Dirac fermions in graphene~\cite{Giovannetti:PRL:12,Ciuti:PRL:12,Pellegrino:PRB:14}.
Additionally, disorder and frustration of the atom-photon
couplings have an important role in the study of BEC in multimode
cavities~\cite{Gopalakrishnan:NatPhys:09,Gopalakrishnan:PRA:10}.
Recent works~\cite{Goldbart:PRL:11,Strack:PRL:11} discussed a
multimodal-Cavity QED simulator with disordered
interactions.
The authors argue that these systems could be employed to
  explore spin-glass properties at the quantum
level~\cite{Goldbart:PRL:11,Strack:PRL:11,Rotondo:PRB:15}. 
However, the possible quantum computation applications of this
new class of quantum simulators remain relatively unexplored.

\begin{figure}
\includegraphics[width=0.3 \textwidth]{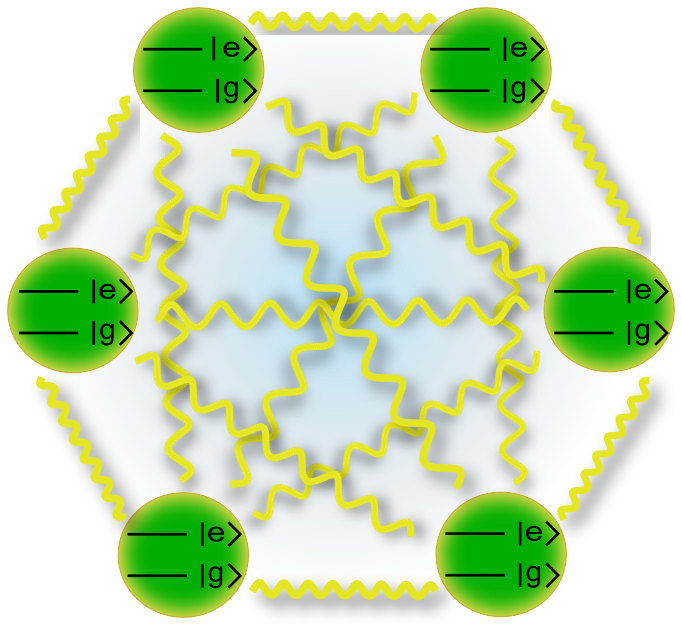}
\caption{In the Dicke model, photons (yellow lines) mediate a long
  range interaction between qubits (green circles). The drawing
  sketches schematically a six qubits system within its
  fully-connected graph and its internal level structure. In the
  standard single-mode Dicke model the exchange coupling is fixed at
  the same value for every pair of qubits. In systems where both many
  modes and disorder are present, the exchange couplings are
  qubit-dependent and take the form given by Eq.~(\ref{coupling}).}
\label{fig:Dicke1}
\end{figure}

In this Letter, we consider a multimode disordered Dicke model with
finite number of modes. We calculate exactly (in the thermodynamic
limit) the free energy of the system at temperature $T = 1/\beta$ and
we find a superradiant phase transition characterized by the same
free-energy landscape of the Hopfield model~\cite{Hopfield:PNAS:82} in
the so-called ``symmetry broken'' phase, with the typical
strong-coupling threshold of the Dicke model. From the
  theoretical standpoint, our results generalize to the case of
  quenched disordered couplings the remarkable analysis performed by
Lieb et al.~\cite{Lieb:AnnPhys:73,Lieb:PRA:73,Wang:PRA:73}. The choice
of frozen couplings is compatible with the characteristic time scales
involved in light-matter interactions. The calculation of the
  partition function leads us to suggest variational ground states
for the model, which we conjecture to be exact in the thermodynamic
limit.
 
The physical consequences of this analysis are fascinating: once the
multimode strong-coupling regime is reached and qubit-photon couplings
are engineered, it should be possible to build a pattern-storing
system whose underlying dynamics is fully governed by quantum
laws. Moreover, Dicke quantum simulators here analyzed may be suitable
to implement specific optimization problems, in the spirit of
  adiabatic quantum
  computation~\cite{Zamponi:12,Farhi20042001,Santoro29032002}. We
point out a non-polynomial optimization problem
\cite{Lucas:FrontPhys:14,Farhi20042001,Santoro29032002}, number
partitioning, which could be implemented in a single mode cavity QED
setup with controllable disorder. Computing applications based on cavity mediated interactions might owns the advantage  to be a viable way to generate entagled many-body states with remarkable scalability properties, as recently shown in Ref. \cite{aron:14}.

Hopfield's main idea~\cite{Hopfield:PNAS:82} is that the retrieval of
stored information, such as memory patterns, may emerge as a
collective dynamical property of microscopic constituents
(``neurons'') whose interconnections (``synapses'') are reinforced or
weakened through a training phase (e.g. \emph{Hebbian
  learning}~\cite{Hebb:ANP:40,Hebb:BML:61}). This is achieved in his
model through a fictitious neuronal-dynamics whose effect is to
minimize the Lyapunov cost function:\begin{equation}
E = -\frac{1}{2} \sum_{i,j=1}^N T_{ij} S_i S_j\,,
\label{Lyapcost}
\end{equation} where $N$ is the number of neurons, $S_i= 1$ if the $i$-th neuron is
active, and $-1$ otherwise, and the $p$ stored patterns
$\xi_i^{(k)}=\pm 1$ ($k = 1,\cdots,p$) determine the interconnections
$T_{ij}$ through the relation: $T_{ij} = 1/N \sum_k
\xi_i^{(k)}\xi_j^{(k)}-p\,\delta_{ij}$.
The analysis in Ref.~\cite{Hopfield:PNAS:82} shows that the long-time
dynamics always converges to one of the $p$ stored patterns,
i.e. these configurations are the global minima of the cost
function~\eqref{Lyapcost}. The interpretation of this result is
that a suitable choice of the interconnections allows to
store a given number of memory patterns into the neural network. Data
retrieval is achieved through an algorithm that minimizes the
energy function~\eqref{Lyapcost}.  A phase transition to a ``complex''
phase marks the intrinsic limitation on the number of patterns $p$
that can be stored. If $p$ exceeds the critical threshold $p\sim 0.14
N$ many failures in the process of retrieval
occur~\cite{Amit:PRA:85,Amit:PRL:85}.

In this manuscript we consider the following multimode Dicke Hamiltonian:
\begin{equation}
H = \sum_{k=1}^M \omega a^{\dagger}_k a^{\phantom'}_k + \Delta \sum_{i=1}^N \sigma^z_i +\sum_{i,k=1}^{N,M} \tilde{g}_{ik} \left(a^{\phantom'}_k + a_k^{\dagger}\right) \sigma^x_i \,,
\label{Dicke}
\end{equation}
effectively modelling quantum light-matter interaction of $N$
two-level systems with detuning $\Delta$ and $M$ electromagnetic modes
supposed to be quasi-degenerate at the common frequency $\omega$ and
with couplings that we parametrize for future convenience as
$\tilde{g}_{ik}=\Omega g_{ik}/\sqrt{N} $, where $\Omega$ is the Rabi
frequency and the dimensionless $g_{ik}$'s are both atom and
mode-dependent. In Cavity QED realizations, $\omega$ represents the
detuning between the cavity frequency and the pumping frequency and
could be both positive or negative. A possible choice of the couplings
is $g_{il}=\cos(k_l x_i)$, being $k_l$ the wave vector of the photon
and $x_i$ the position of $i$-th atom~\cite{Strack:PRL:11}.

We are interested in the thermodynamic properties of this system in
the limit $M \ll N$, and thus in evaluating the partition function $Z
= \Tr e^{-\beta H}$. This evaluation can be performed rigorously in
the thermodynamic limit ($N\rightarrow \infty$) using the techniques
introduced in Refs.~\cite{Lieb:AnnPhys:73,Lieb:PRA:73,Wang:PRA:73}. We
first consider the fully-commuting limit $\Delta = 0$. In this case
the evaluation of the partition function is straightforward~(see Supplementary material)
and we obtain $Z = Z_{FB}\, Z_H$, where $Z_{FB}$ is a free boson
partition function and $Z_H$ is a classical Ising model with local
quenched exchange interactions of the form:
\begin{equation}
J_{ij} = -\frac{\Omega^2}{N} \sum_{k=1}^M \frac{g_{ik}\, g_{jk}}{\omega}\,.
\label{coupling}
\end{equation}  
The physical interpretation of this result is that photons
mediate long range interactions among the atoms, resulting in an
atomic effective Hamiltonian described by a fully-connected Ising
model (see Fig. \ref{fig:Dicke1}). The role of the couplings $g_{ik}$
can be understood from Eq.~\eqref{Lyapcost} in the context of the
Hopfield network. They are the memory pattern stored in the system. By
computing exactly the free energy of the model, we will show
that this interpretation stays unaltered in the more complicated case
$\Delta \neq 0$.

We now proceed to the evaluation of the quantum partition function. We
use the method of Wang and
coworkers~\cite{Wang:PRA:73,Lewenstein:NJP:09} (proved to be exact in
the thermodynamic limit for $ M/N \rightarrow 0$~\cite{Lieb:PRA:73}).
We introduce a set of coherent states $\ket{\alpha_k}$ with
$\alpha_k=x_k+iy_k$, one for each electromagnetic mode $k$, and we
expand the partition function on this overcomplete basis:
\begin{equation}
Z = \int \prod_{k=1}^M \frac{d^2 \alpha_k}{\pi}\, \Tr_A \bra{\{\alpha\}} e^{-\beta H} \ket{\{\alpha\}}\,,
\label{pf}
\end{equation}
where $\Tr_A$ is the atomic trace only. 
The only technical
  complication is the calculation of the matrix element
in~\eqref{pf}. This turns out to be equal, apart from non-extensive
contributions, to the exponential of the operator in Eq~\eqref{Dicke}
with the replacements $a_k,~a^{\dagger}_k \rightarrow
\alpha_k,~\alpha_k^\ast$ ~\cite{Wang:PRA:73,Lieb:PRA:73}. At this
stage the trace over the atomic degrees of freedom can be easily
performed. The integral over the imaginary parts of $\alpha_k$'s 
give an overall unimportant constant. Finally, defining the
$M$-dimensional vectors $\mathbf{x}=(x_1,x_2,\cdots, x_M)$ and
$\mathbf{g}_i= (g_{i1},g_{i2},\cdots, g_{iM})$, and with the change of
variables $\mathbf{m} = \mathbf{x}/\sqrt N$, the partition function
assumes a suitable form for performing a saddle-point integration,
i.e. $Z = \int \, d^M \mathbf{m}\, e^{-N f(\mathbf{m})}$. Here $f$ is
the free energy
\begin{equation}
f(\mathbf{m}) = \beta \mathbf{m} \cdot \mathbf{m} - 
\frac{1}{N} \sum_{i=1}^N \log G(\mathbf{m},\mathbf{g}_i)\, , 
\label{MF_free}
\end{equation}
with:  $
G(\mathbf{m},\mathbf{g}_i) = 2 \cosh \left[ \beta \left(\Delta^2+ \Omega^2 (\mathbf{g}_i \cdot \mathbf{m})^2\right)^{\frac{1}{2}}\right] 
$.

The order parameter $\mathbf{m}$ describes the superradiant phase
transition. Physically, it gives the mean number of photons in every
mode~\cite{Emary:PRL:03}. Its value is determined by minimizing the
free energy in Eq.~ (\ref{MF_free}).  Solutions of this optimization
problem are, in principle, $\mathbf{g}_i$-dependent, but in the
thermodynamic limit both the free energy and the saddle-point equation
are self-averaging~\cite{Amit:PRA:85}. Thus we conclude that the free
energy and the saddle point equations are given by
\begin{align}
 f(\mathbf{m}) &= \beta \mathbf{m} \cdot \mathbf{m} - \Braket{\log G(\mathbf{m},\mathbf{g})}_{\mathbf{g}}\,,\nonumber\\
\mathbf{m} &= \frac{\Omega^2}{2} \Braket{\frac{(\mathbf{g}\cdot\mathbf{m})\, \mathbf{g}}{\mu(\mathbf{g})}\tanh \left(\beta \mu(\mathbf{g})\right)}_{\mathbf{g}}\,, 
\label{saddle}
\end{align}
with: $\mu(\mathbf{g}) = \left(\Delta^2+ \Omega^2 (\mathbf{g} \cdot
  \mathbf{m})^2\right)^{\frac{1}{2}}$ and
$\braket{\cdots}_{\mathbf{g}}$ representing the average over the
disorder distribution.
Eq.~\eqref{saddle} reduces to the mean-field equations for the
Hopfield model for $\Delta \rightarrow 0$
\cite{Amit:PRA:85}. Thus, $\Delta$ may be intended as a
  quantum annealer parameter. To fully specify the model, the
probability distribution for the couplings is needed. In the following
we will assume
\begin{equation}
P(\mathbf{g}) = \prod_{k=1}^M \left(\frac{1}{2}\delta \left(g_k - 1\right) + \frac{1}{2}\delta \left(g_k + 1\right)\right)\,,
\end{equation}
but we have verified that the results are qualitatively robust as
  long as the disorder is not too peaked around zero 
  in accordance with the classical results of Ref.~\cite{Amit:PRA:85} 
 To locate the critical point
it suffices to expand in Taylor series Eqs.~(\ref{saddle}). As in the conventional Dicke model, a temperature-independent
threshold $\Omega_c^2 = 2 \Delta$ emerges. For $\Omega < \Omega_c$,
the phase transition is inhibited at all temperatures. Whenever the
magnitude of the coupling exceeds this threshold value, the
critical temperature is located at $T_c = \Delta/\arctanh
\left(2\Delta/\Omega^2\right)$.

\begin{figure}
\includegraphics[width=0.42 \textwidth]{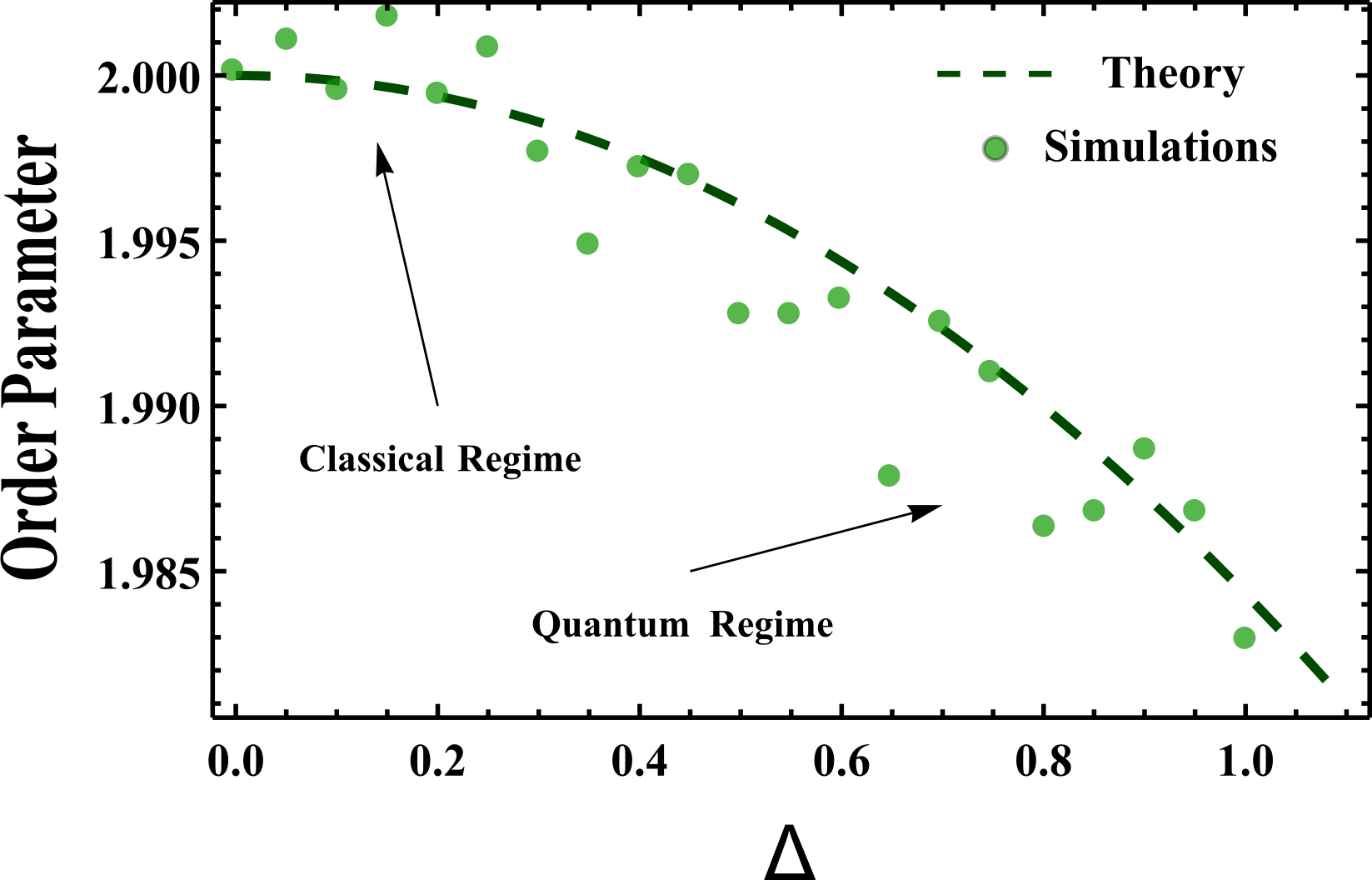}
\caption{Order parameter $m^{(1)}$ in Eq. (\ref{m1t0}) as a function
  of $\Delta$ in the ultrastrong coupling regime at $T=0$. On the
  theoretical curve (dashed line) are superposed the results (green
  dots) of a naive numerical optimization algorithm minimizing the
  ground state energy ansatz in Eq. (\ref{ansatz}). The parameters of
  the simulation are $\Omega = 4$, $N=100$, $M=10$. Green dots are the
  result of a single disorder realization. This implies the
  correctness of our self-averaging hypothesis, which is a good
  approximation already for $N=100$. The classical Hopfield model is
  recovered in the limit $\Delta \rightarrow 0$.}
\label{fig:Dicke2}
\end{figure}

Above the critical temperature $T_c$ the only solution to
  (\ref{saddle}) is a paramagnetic state, with $m_k = 0$ for all
  $k$. Below $T_c$, different solutions appear. We now set out to
  classify these solutions and their stability under temperature
  decrease. For this analysis, we both considered the Hessian matrix
  $\partial^2{f}/\partial{m_k}\partial{m_l}$ (see Supplementary
  materials for its explicit expression) and numerical optimization
  (Figure \ref{fig:Dicke2}) . The key point, as mentioned above is
  that in this ``symmetry-broken'' phase the system takes $2M$
  degenerate ground states (as well as many metastable states
  energetically well separated from the ground states). In other
  words, also in this fully quantum limit the free-energy landscape
  still closely resembles that of the Hopfield
  model~\cite{Amit:PRA:85}.

The ground state solutions have the explicit form:
\begin{equation}
\mathbf{m}_k = m^{(1)}(\underbrace{0,0,\cdots,0}_{k-1 \,\,times},\underbrace{1,0,\cdots,0}_{M-k+1 \,\,times})\,. 
\end{equation}
Equation (\ref{saddle}) for the order
parameter $m^{(1)}$ reduces to: $2 \mu(m^{(1)}) = \Omega^2 \tanh\left(\beta \mu(m^{(1)})\right)$, where $\mu(m^{(1)}) = \sqrt{\Delta^2 + \Omega^2 (m^{(1)})^2}$. In the zero temperature limit the order parameter can be evaluated exactly:
\begin{equation}
m^{(1)} = \pm\sqrt{\frac{\Omega^2}{4} - \frac{\Delta^2}{\Omega^2}}\,.
\label{m1t0}
\end{equation}

At zero temperature the most interesting
state is the ground state (GS) of the Hamiltonian
(\ref{Dicke}). Inspired by the calculation above we propose the
variational ansatz for the GS:
\begin{equation}
  \ket{GS} = \ket{\alpha_1,\alpha_2,\cdots,\alpha_M} \ket{\text{spin}(\alpha_1,\cdots,\alpha_M)}\,,
\end{equation}
where $\ket{\alpha_1,\alpha_2,\cdots,\alpha_M}$ is the product of $M$
coherent states and the spin part is factorized. The mean value of the
energy in this GS is given by:
\begin{equation}
E_{GS}(\mathbf{m}) = \mathbf{m} \cdot \mathbf{m} - \Braket{\sqrt{\Delta^2 + \Omega^2 (\mathbf{g}\cdot \mathbf{m})^2}}_{\mathbf{g}}\,.
\label{ansatz}
\end{equation}
This expression exactly equals the free-energy computed previously in
the limit $\beta \rightarrow \infty$, which leads us to
conjecture that our factorized variational ansatz is exact.  The quantum
phase transition is located at the critical coupling $\Omega =
\Omega_c=\sqrt{2\Delta}$, at which the paramagnetic solution becomes
unstable.  In the symmetry-broken phase we have $2M$ degenerate ground
states of the form
\begin{equation}
\ket{GS}_k = \ket{\underbrace{0,0,\cdots,0}_{k-1 \,times},\pm m^{(1)},\underbrace{0,\cdots,0}_{M-k\,times} }\ket{\text{spin}(\pm m^{(1)})}_k\,,
\label{GS}
\end{equation}
with $k = 1,\cdots M$. The spin wave function is also factorized
$\ket{\text{spin}(\pm m^{(1)})}_k = \prod_i \ket{s_i}_k$, with
\begin{equation}\label{eq:si}
\ket{s_i}_k = \frac{1}{\mathcal{N}}\left(-\frac{-\Delta +\sqrt{\Delta^2 + \beta_{ik}^2} }{\beta_{ik}}\ket{e_i} + \ket{g_i}\right)\,, 
\end{equation}
where $\mathcal{N}$ a normalization, $\beta_{ik} = g_{ik} m^{(1)}$
and $\ket{e_i}$,
$\ket{g_i}$ are $\sigma_i^z$'s eigenstates. 
It is worth noting that, as expected,
the ground state energy is a self-averaging quantity, whereas the
ground states are not, being disorder-dependent also in the
thermodynamic limit.

The above calculation shows that
in the superradiant phase the ground state of the system is a quantum
superposition of the $2M$ degenerate eigenvectors given by
Eqs.~(\ref{GS},\ref{eq:si}). Their explicit form suggests that at fixed disorder
and mode number the information about the disordered couplings
belonging to the $k$-th mode is \emph{printed} on the atomic wave
function. Moreover, the photonic parts of the wave functions are all
orthogonal for $k_1 \neq k_2$ in the thermodynamic limit. This
implies that in principle a suitable measure on the photons-subsystem
causes the collapse over one of the $2M$ ground states and gives thus
the possibility to retrieve information (``patterns'') stored in the
atomic wave function. As mentioned above, a single-mode Dicke model
has been recently realized with cavity-mediated Raman transitions in
cavity QED with ultracold atoms \cite{Baden:PRL:14}. A Multimode
cavity QED setup supporting disordered couplings has been proposed in
refs.~\cite{Goldbart:PRL:11,Strack:PRL:11}, and preliminary evidence
of superradiance in this system is found
  in~\cite{Renzoni:PRA:13}. A setup operating in multimode regime has
been recently suggested also in Circuit QED~\cite{Egger:PRL:13}. We
are not aware of a setup (in Cavity or Circuit QED) that implements
both multimode strong coupling regime and controllable disorder, a
necessary condition for the quantum pattern-retrieval system that we
conjecture here.

We surmise that Multimode Dicke quantum setups with controllable
disorder could be used beyond storage, to simulate specific
optimization problems. Indeed, finding the ground state of
  classical spin models with disordered interactions is equivalent, in
  most cases, to finding solutions of computationally expensive
  non-polynomial (NP) problems~\cite{Lucas:FrontPhys:14}. For
example, the \emph{simplest} NP-hard problem, number partitioning,
could be implemented in a single-mode cavity QED setup with
controllable disorder as follows. Number partitioning can be
formulated as an optimization problem~\cite{Mertens:PRL:98}: given a
set $\mathcal{A} = \{a_1,a_2,\dots,a_N\}$ of positive numbers, find a
partition, i.e. a subset $\mathcal{A}' \subset \mathcal{A}$, such that
the residue: $E = |\sum_{a_j \in \mathcal{A}'} a_j - \sum_{a_j \notin \mathcal{A}'} a_j\, |$ is minimized. A partition can be defined by numbers $S_j = \pm 1$:
$S_j = 1$ if $a_j \in \mathcal{A}'$, $S_j = -1$ otherwise. The cost
function can be replaced by a classical spin hamiltonian, whose ground state is
equivalent to the minimum partition:
\begin{equation}
H = \sum_{i,j=1}^N a_i a_j S_i S_j\,.
\end{equation}   
In a single mode cavity QED network couplings can be chosen as $g_i =
\cos(k x_i)$ \cite{Strack:PRL:11}. By the definition of $a =~
\max_{\mathcal{A}} a_j$ and $\tilde{a}_j = a_j/a$, it is possible to
engineer the $g_i$'s in such a way to implement a given instance of
the problem provided that the cavity is in the ``blue'' detuned regime
to ensure the appropriate sign for the couplings, see
Eq. (\ref{coupling}). With a suitable adiabatic annealing of
  the atomic detuning $\Delta$, the system should collapse on qubit
  configurations that are good solutions of the corresponding
  optimization problem.

In conclusion, this Letter provides the first rigorous analysis
of the multimode disordered Dicke model, valid beyond the
weak-coupling regime and exact in the thermodynamic limit. The
  equivalence between multimodal disordered Dicke model and a quantum Hopfield
  network~\cite{Nonomura:95}, together with the proposal of a cavity QED setup implementing a non polynomial optimization problem, demonstrates the possibility of quantum
  computational abilities of this new class of quantum simulators.  Our proposal is
conceptually complementary to a standard quantum computation
perspective~\cite{Loss:PRA:98,Raussendorf:PRA:03}. Indeed, the
  information can be ``written'' on the qubits through a quantum
  annealing on the detuning $\Delta$, similarly to what happens for
  adiabatic quantum
  computation~\cite{Farhi20042001,Santoro29032002,Zamponi:12}.

\emph{Acknowledgements}.--- We are grateful to B.~Bassetti,
S.~Mandr\`a, G.~Catelani, M.~Gherardi, S.~Caracciolo, L.~Molinari,
F.~Solgun and A.~Morales for useful discussions and feedback on this
manuscript. GV was supported by Alexander von Humboldt foundation and
Knut och Alice Wallenbergs foundation.

\bibliographystyle{apsrev4-1}

\bibliography{PietroBibliography2}

\onecolumngrid
\newpage

\renewcommand{\thefigure}{S\arabic{figure}}
 \setcounter{figure}{0}
\renewcommand{\theequation}{S.\arabic{equation}}
 \setcounter{equation}{0}
 \renewcommand{\thesection}{S.\Roman{section}}
\setcounter{section}{0}

\section*{Dicke simulators with emergent collective quantum computational abilities\\ supporting material}
\subsection*{Pietro Rotondo, Marco Cosentino Lagomarsino, and Giovanni Viola }

In this material, we give more details on the derivations of the results presented in the main text. 

\section{Derivation of Equation (3)}
We begin with the derivation of Eq. (3), of the main text. We consider the partition function $Z=\mathrm{Tr}e^{-\beta H}$ with $H$ given in Eq. (2) for  $\Delta = 0$.  In this fully commuting limit we can evaluate the partition function straightforwardly. 
We introduce new set of bosonic operators: 
\begin{equation}
b^{\dagger}_k = a^{\dagger}_k + \frac{\Omega}{\omega\sqrt N}\sum_{i=1}^N g_{ik} \sigma^x_i\,,   \qquad  b_k = a_k + \frac{\Omega}{\omega\sqrt N}\sum_{i=1}^N g_{ik} \sigma^x_i \,.
\end{equation}  
with $[b_{k'},b^{\dagger}_k]=\delta_{k,k'}$. By means of those, $H$ is written as the sum of two commuting operators:
\begin{equation}
H = \omega \sum_{k=1}^M b^{\dagger}_k b^{\phantom{\dagger}}_k - \frac{\Omega^2}{N \omega} \sum_{i,j=1}^N \sum_{k=1}^M g_{ik}g_{jk} \sigma^x_i \sigma^x_j 
\end{equation}   
As a byproduct we obtain the factorization of the full partition function \cite{Rotondo:PRB:15s}:$
 Z = Z_B \, Z_{H} 
$, where $Z_B$ is an overall free boson partition function that we can safely ignore in the thermodynamic limit. On the other hand
\begin{equation}\label{eq:sup:ZF}
 Z_H=\mathrm{Tr}_{\sigma}\exp\left(\beta\sum_{i,j}J_{ij} \sigma_i \sigma_j \right)
 \end{equation} 
 is an Ising contribution with both spin and mode dependent couplings $J_{ij} $ of the form given in Eq. (3) of the main text.
  In Eq.\eqref{eq:sup:ZF}  $\mathrm{Tr}_{\sigma}$ indicates the trace on the spins only.

\section{Derivation of Equation (5)}
In this section we report the derivation of Eq. (5) of the main text, which essentially follows the derivation of Wang and Hioe~\cite{Wang:PRA:73s} proved to be rigorous by Hepp and Lieb \cite{Lieb:PRA:73s}. 
The authors of Ref.~\cite{Wang:PRA:73s}  have shown explicitly, in the termodynamic limit, that the convenient way to calculate the trace on the Hilbert space of  bosons, in the partition function, is to evaluate it on the set of the coherent states $\vert \{\alpha\} \rangle$.  
The photonic matrix element in the partition function of Eq. (4) equals in the thermodynamic limit ($\alpha_k = x_k + i y_k$):
\begin{equation}
\bra{\{\alpha\}} e^{-\beta H} \ket{\{\alpha\}} \simeq \exp\left(-\beta \omega \sum_{k=1}^M (x_k^2+y_k^2) - \beta \Delta \sum_{i=1}^N \sigma^z_i - \beta\frac{\Omega}{\sqrt N}\sum_{i=1}^N \sum_{k=1}^M g_{ik} x_k \sigma^x_i \right)\,.
\end{equation}
The atomic trace thus factorizes and  it can be calculated: 
\begin{equation}
Z = \int \prod_{k=1}^M \,\frac{dx_k\,dy_k}{\pi} \, e^{-\beta \omega \sum_{k=1}^M (x_k^2+y_k^2)} \prod_{i=1}^N \cosh \left(\beta \sqrt{\Delta^2 + \frac{\Omega^2}{N} \left(\sum_{k=1}^M g_{ik}x_k\right)^2}\right) =  \int \prod_{k=1}^M \,\frac{dm_k}{\pi} e^{-N f(\mathbf{m},\beta)}\,.
\end{equation} 
In the last term we introduced the vectorial notation defined in the main text.  The final expression for the free energy is:
\begin{equation}
f(\mathbf{m},\beta) = \beta \mathbf{m} \cdot \mathbf{m} - 
\frac{1}{N} \sum_{i=1}^N \log \cosh\left(\beta \sqrt{\Delta^2 + \frac{\Omega^2}{N} \left(\mathbf{g}_i \cdot \mathbf m\right)^2}\right)\,, 
\label{freeenergy}
\end{equation}
By minimizing the free energy above and using the self averaging property of Eq. (\ref{freeenergy}), we obtain the exact mean field equations:
\begin{equation}
\mathbf{m} = \frac{\Omega^2}{2} \Braket{\frac{(\mathbf{g}\cdot\mathbf{m})\, \mathbf{g}}{\mu(\mathbf{g})}\tanh \left(\beta \mu(\mathbf{g})\right)}_{\mathbf{g}}
\label{meanfield}
\end{equation}
To locate the critical point it suffices to expand in Taylor series Eqs.~(\ref{freeenergy}),(\ref{meanfield}):
\begin{align}
 f(\mathbf{m})- f(\mathbf{0}) &= \beta\left(1- \frac{\Omega^2}{2\Delta} \tanh (\beta \Delta)\right) \mathbf{m} \cdot \mathbf{m} + O\left(m_k^4\right)\,,\qquad
 m_k = \frac{\Omega^2}{2\Delta} \tanh (\beta \Delta)\,m_k + O(m_k^3)\,.
\label{saddle_tay}
\end{align}
As in the conventional Dicke model, a temperature-independent
threshold $\Omega_c^2 = 2 \Delta$ emerges. For $\Omega < \Omega_c$,
the phase transition is inhibited at all temperatures. Whenever the
magnitude of the coupling exceeds this threshold value, the
critical temperature is located at $T_c = \Delta/\arctanh
\left(2\Delta/\Omega^2\right)$.
Solutions to Eq. (\ref{meanfield}) can be classified according to
the number of non-zero components $n$ of the order parameter
$\mathbf{m}$ \cite{Amit:PRA:85s}:
\begin{equation}
\mathbf{m}_n = m^{(n)}(\underbrace{1,1,\cdots,1}_{n \,\,times},\underbrace{0,0,\cdots,0}_{M-n \,\,times})\,, 
\end{equation}
where all permutations are also possible. In particular, solutions for
$n=1$ are the ones with the lowest free energy. There are $2M$ of such
solutions, corresponding to the $\mathbb{Z}_2 \times S_M$ symmetry
breaking of our multimode Dicke model. 

For completeness we report the explicit expression for the Hessian matrix of the free energy, omitted in the main text for space imitations:
\begin{align}&
\frac{\partial^2{f}}{\partial{m_k} \partial{m_l}} = 
2 \delta_{kl} - \Omega^2 \Braket{\frac{g_k g_l}{\mu(\mathbf{g})}\tanh \left(\beta\mu(\mathbf{g})\right)}_{\mathbf{g}} 
+\Omega^4  \Braket{\frac{(\mathbf{g} \cdot \mathbf{m})^2\,g_k g_l}{\mu(\mathbf{g})^2}\left[\frac{\tanh \left(\beta\mu(\mathbf{g})\right)}{\mu(\mathbf{g})}+\beta \mathrm{sech}^2\left(\beta\mu(\mathbf{g})\right)\right]}_{\mathbf{g}}\,.  
\end{align}

\end{document}